\documentclass[12pt]{article}
\title{Finiteness Properties of the N=4 Super-Yang--Mills~Theory in
 Supersymmetric Gauge}
\usepackage{mathrsfs}
\usepackage{amsmath}

\global\arraycolsep=1pt
\usepackage[colorlinks=true,backref=true,linkcolor=black,anchorcolor=black,citecolor=black,filecolor=black,menucolor=black,pagecolor=black,urlcolor=black]{hyperref}
\usepackage[Symbol]{upgreek}
\usepackage{bbm}
\usepackage{dsfont}
\usepackage{amssymb}
\usepackage{textcomp}

\newcommand{\Scal}[1]{\biggl ({#1} \biggr )}
\newcommand{\scal}[1]{\bigl ({#1} \bigr )}
\def\bea{\begin{eqnarray}} 
\def\eea{\end{eqnarray}}
\def\be{\begin{equation}}
\def\ee{\end{equation}}
\newcommand{\CR}{\nonumber \\*}

\newcommand{\trace}{\hbox {Tr}~}

\newcommand{\gra}[2]{{\scriptscriptstyle (#1 , #2 )}}
\newcommand{\ord}[1]{{\scriptscriptstyle (#1)}}
\def\L{{\cal L}}

\DeclareMathAlphabet{\mathpzc}{OT1}{pzc}{m}{it}
\def\s{\,\mathpzc{s}\,}
\def\a{{\scriptscriptstyle (\mathpzc{s})}}
\def\q{{\scriptscriptstyle (Q)}}
\def\qs{{\scriptscriptstyle (Q\mathpzc{s})}}

\def\Lc{\mathscr{L}}

\def\Q{{\mathcal{S}_{\q}}}
\def\phiq{{\varphi^{\q}}}
\def\phiqs{{\varphi^{{\qs}}}}
\def\Omegas{{\Omega^\a}}
\def\Omegaq{\Omega^{\q}}
\def\Omegaqs{\Omega^{{\qs}}}

\def\cq{c^{{\q}}}
\def\muq{\mu^{{\q}}}
\def\A{a}

\def\C{c}

\def\phis{{\varphi^\a}}
\def\chis{{\upchi^{\a}}}
\def\chiq{{\upchi^{\q}}}
\def\chiqs{{\upchi^{\qs}}}

\def\S{{\mathcal{S}_\a}}
\def\P{\mathcal{P}}
\def\F{\mathscr{F}}

\def\aGg{{\mathcal{G}^\bullet}}
\def\Gg{\overline{{\mathcal{G}_\bullet}}}

\def\LaGg{L\mathcal{G}^\bullet}
\def\LGg{\overline{{L\mathcal{G}_\bullet}}}

\newcommand{\ins}[1]{\S_{|\Gamma} \bigl[
 \Uppsi^{\scriptscriptstyle{(#1)}} \cdot \Gamma \bigr]}

\def\C{\mathcal{C}}

\def\R{{\scriptscriptstyle \mathcal{R}}}
\def\rp{{\mathpzc{m}}}
\def\p{{\scriptscriptstyle \rm inv}}
\usepackage[colorlinks=true,backref=true,linkcolor=black,anchorcolor=black,citecolor=black,filecolor=black,menucolor=black,pagecolor=black,urlcolor=black]{hyperref}

\usepackage[Symbol]{upgreek}
\usepackage{caption}
\usepackage{bbm}
\usepackage{dsfont}
\usepackage{amssymb}
\usepackage{textcomp}
\def\bea{\begin{eqnarray}}
\def\eea{\end{eqnarray}}
\def\be{\begin{equation}}
\def\ee{\end{equation}}

\def\L{{\cal L}}

\def\Lc{\mathscr{L}}

\def\bomega{{\overset{\circ}{\omega}}}

\def\susy{{\delta^{\mathpzc{Susy}}}}

\def\bomega{\varpi}

\begin{document}
\allowdisplaybreaks[1]
\renewcommand{\thefootnote}{\fnsymbol{footnote}}

\begin{titlepage}
%\null
\begin{flushright} CERN-PH-TH/2006-090\\
 \end{flushright}
\begin{center}
 {{\Large \bf
 Finiteness Properties of the $\mathcal{N}=4$ Super-Yang--Mills~Theory in
 Supersymmetric Gauge}}
\lineskip .75em
\vskip 3em
\normalsize
{\large L. Baulieu\footnote{email address: baulieu@lpthe.jussieu.fr},
 G. Bossard\footnote{email address: bossard@lpthe.jussieu.fr},
 S. P. Sorella\footnote{email address: sorella@uerj.br}}\\
 $^{* }$\it Theory Division, CERN, Switzerland\footnote{
1211-Geneve 23, Switzerland}
 \\
$^{*\dagger}$ {\it LPTHE, CNRS and Universit\'es Paris VI - Paris VII, Paris,
France}\footnote{
4 place Jussieu, F-75252 Paris Cedex 05, France.}
\\
$^{\ddagger}${\it Departamento de F\'\i sica Te\'orica
Instituto de F\'\i sica, UERJ, Universidade do Estado do Rio de
Janeiro, Brazil \footnote{Rua S\~ao Francisco Xavier 524, 20550-013 Maracan\~a
Rio de Janeiro, Brazil.}}

\vskip 1 em
\end{center}
\vskip 1 em
\begin{abstract}
With the introduction of shadow fields, we demonstrate the
renormalizability of the $\mathcal{N}=4$ super-Yang--Mills theory
in component formalism, independently of the choice of UV
regularization. Remarkably, by using twisted representations, one
finds that the structure of the theory and its renormalization is
determined by a subalgebra of supersymmetry that closes off-shell.
Starting from this subalgebra of symmetry, we prove some features
of the superconformal invariance of the theory. We give a new
algebraic proof of the cancellation of the $\beta$ function and we
show the ultraviolet finiteness of the $1/2$ BPS operators at all
orders in perturbation theory. In fact, using the shadow field as a
Maurer--Cartan form, the invariant 
polynomials in the scalar fields in traceless symmetric
representations of the internal R-symmetry group are simply
related to characteristic classes. Their UV finiteness is a
 consequence of the Chern--Simons formula.
\end{abstract}

\end{titlepage}
%%%%%%%%%%%%%%%%%
\renewcommand{\thefootnote}{\arabic{footnote}}
\setcounter{footnote}{0}

%%%%%%%%%%%%%%%%%%%%%%%%%%%%%%%%%%%%%%%%%%%%%%%%%%%%%%%%

%%%%%%%%%%%%%%%%%%%%%%%%%%%%%%%%%%%%%%%%%%%%%%%%%%%%

\renewcommand{\thefootnote}{\arabic{footnote}}
\setcounter{footnote}{0}

%%%%%%%%%%%%%%%%%%%%%%%%%%%%%%%%%%%%%%%%%%%%%%%%%%%%%%%%

%\tableofcontents

%%%%%%%%%%%%%%%%%%%%%%%%%%%%%%%%%%%%%%%%%%%%%%%%%%%%%%%%

\section{Introduction}
In a recent paper, new fields have been introduced for
supersymmetric gauge theories, which we called shadow fields.
These fields are elements of BRST doublets. They determine a
BRST like operator for the supersymmetry invariance,
which is fully compatible with the gauge symmetry \cite{shadow}.
Shadow fields also allow for the construction of new classes of
gauges that interpolate between the usual Faddeev--Popov gauges
and new ones, which are explicitly supersymmetric. These gauges
give an additional Slavnov--Taylor identity, for controlling
supersymmetry at the quantum level. The observables are determined
from the cohomology of the BRST operator for gauge
invariance, and their supersymmetry covariance can be established
from the new Slavnov--Taylor identity.

It has been conjectured for many years that the $\mathcal{N}=4$
super-Yang--Mills theory is a superconformal theory. Its $\beta$
function vanishes at all orders in perturbation theory
\cite{beta,conform}. The superconformal invariance is a basic
feature of the $AdS / CFT$ Maldacena's conjecture
\cite{maldacena}. In this paper we use the new supersymmetric
gauges to directly prove in perturbation theory
 some of the results that are implied by the
conformal invariance of the $\mathcal{N}=4$ super-Yang--Mills theory,
with the only symmetry hypothesis of super Poincar\'e supersymmetry.

% We only suppose that the
%theory is renormalized in such way to preserve both gauge invariance
%and $\mathcal{N}=4$ supersymmetry.

In \cite{shadow}, the renormalization of supersymmetric theories,
using shadow fields, was detailed when the supersymmetry algebra
closes off-shell, that
 is, for cases where auxiliary fields exist for closing the whole
 supersymmetry algebra. In the $\mathcal{N}=4$
 super-Yang--Mills theory, no such auxiliary fields exist.
Here, we will bypass this point by combining
 the methodology of \cite{shadow} and the existence of a
 supersymmetry
 subalgebra with 9 generators, which is large
 enough to constrain the classical $\mathcal{N}=4$ super-Yang--Mills
 action, and small enough to be closed without the use of equations
of motion. This subalgebra was introduced in \cite{BBT,N4}, using
methods that are specific to topological field theory and involve
twisted variables.

The Slavnov--Taylor identities that are allowed by the
introduction of the shadow fields enable us to prove the
renormalizability of the $\mathcal{N}=4$ theory, without any
assumption on the choice of the ultraviolet regularization. In this paper,
we show the absence of a possible anomaly for the $\mathcal{N}=4$ theory.
 Algebraic methods have been used to show
 that invariant polynomials depending on one of
the scalar fields of the supersymmetry multiplet are protected from
renormalization \cite{Aless}.
We will extend this result to the whole set of local observables of the 
$\mathcal{N}=4$ super-Yang--Mills theory, which are invariant polynomials
in scalar fields, taking their values in any given traceless symmetric
representation of the R-symmetry group. These operators are the $1/2$
BPS primary operators. We then obtain the finiteness
of the whole $1/2$ BPS multiplets, by supersymmetry covariance. We will give a
new proof of the cancellation of the $\beta$ function at all
orders in perturbation theory. Remarkably, all these results are
obtained by using small sectors of the supersymmetry algebra. The
latter exist, thanks to the possibility of twisting the
supersymmetry algebra, by combining the R-symmetry and the Lorentz
symmetry. In order to prove the stability of the action under
renormalization, it is in fact sufficient to use a sector of the
supersymmetry algebra with 6 generators. And, to prove that all $1/2$
BPS primary operators are protected operators, one uses a sector
of the supersymmetry algebra with 5 generators. A key feature
of this proof is the Chern--Simons formula, where the shadow
field can be identified as a Maurer--Cartan form. It appears
that the scalar observables of topological field theories
determine the $1/2$ BPS primary operators by covariance under
the R-symmetry of the supersymmetric theory. In fact, their finiteness property is closely related to the existence of characteristic classes.

We will suppose everywhere that the gauge group is simple.
%%%%%%%%%%%%%%%%%%%%%%%%%%%%%%%%%%%%%%%%%%%%%%%%%%%%%%%%
\section{$\mathcal{N}=4$ super-Yang--Mills theory in the twisted variables}
\subsection{Fields and symmetries}

Consider the $\mathcal{N}=4$ multiplet $(A,\lambda^\alpha,\phi^i)$ in a
flat euclidean space $Spin(4) \cong SU(2)_+\times SU(2)_- $, where
$\alpha, i$ are indices in the $\bf 4$ and the $\bf 6$ of the internal symmetry $SL(2,\mathds{H})$, which
is the euclidean version of the $SU(4)$ R-symmetry in Minkowski
space. The components of spinor and scalar fields
$\lambda^\alpha$ and $\phi^i$ can be twisted, i.e., decomposed on
irreducible representations of the following
subgroup\footnote{Usually, one means by twist a redefinition of
 the energy momentum tensor that we do not consider here.}
\be
SU(2)_+\times {\rm diag} \scal{SU(2)_-\times SU(2)_R} \times U(1)
\subset SU(2)_+\times SU(2)_-\times SL(2,\mathds{H})
\ee
We redefine $SU(2) \cong {\rm diag}\scal{SU(2)_-\times
 SU(2)_R}$. The $\mathcal{N}=4$ multiplet is decomposed as
 follows
\be\label{dec}
( A_\mu, \Psi_\mu, \eta, \chi^I, \Phi, \bar \Phi)
\hspace {10mm}
(L, h_I, \bar \Psi_\mu, \bar \eta, \bar \chi_I)
 \ee
In this equation, the vector index $\mu$ is a ``twisted world
index'', which stands for the $(\frac{1}{2}, \frac{1}{2})$
representation of $ SU(2)_+\times SU(2)$. The index $I$ is for the
adjoint representation of the diagonal $SU(2)$. In fact, any given
field $X^I$ can be identified as a twisted antiselfdual $2$-form
$X_{\mu\nu^-}$, by using the flat hyperK\"{a}hler structure $J^I_{\mu\nu}$.

The twisted ``matter'' multiplet $ (L, h_I, \bar \Psi_\mu, \bar
\eta, \bar \chi_I) $ involves therefore four scalars, which are now
assembled as a scalar $L$ and an antiselfdual $2$-form $h_I$. For the
sake of clarity, we will shortly display a table with the
representations of the twisted fields under the various symmetries, as
well as their commutation properties.

The ten-dimensional super-Yang--Mills theory determines by dimensional
reduction the untwisted $\mathcal{N}=4$ super-Yang--Mills
theory. Analogously, 
 the twisted eight-dimensional $\mathcal{N}=2$ theory determines 
the twisted formulation of the $\mathcal{N}=4$
super-Yang--Mills theory in four dimensions \cite{BKS}. An horizontality conditions is given in \cite{BBT}, which
determines the maximal supersymmetry subalgebra of the twisted $\mathcal{N}=2,\ d=8 $ theory,
which can be closed without the equations of motion. We will consider
this horizontality condition with minor modifications, which are
convenient for perturbation theory. It defines the BRST operator
$\s$ associated to gauge invariance and the graded differential
operator $Q$, which generates the maximal off-shell closed
supersymmetry subalgebra. The latter depends on nine twisted
supersymmetry parameters, which are one scalar $\bomega$ and one
eight-dimensional
vector $\varepsilon$. $\s$ and $Q$ are thus obtained by expanding
over all possible gradings the following horizontality condition
\begin{gather}
\label{extcurvature}
( d + \s + Q - \bomega i_\varepsilon ) \scal{ A + \Omega + c } + \scal{
 A + \Omega + c}^2 \hspace{40mm}\CR
\hspace{50mm} = F + \bomega \Psi + g(\varepsilon) \eta + i_{\varepsilon}
\chi + \bomega^2 \Phi + |\varepsilon|^2 \bar\Phi
\end{gather}
and its associated Bianchi relation
\begin{gather}\label{bian}
( d + \s + Q - \bomega i_{\varepsilon} ) \scal{F + \bomega \Psi +
 g(\varepsilon) \eta + i_{\varepsilon} \chi + \bomega^2 \Phi +
 |\varepsilon|^2 \bar\Phi}
\hspace{30mm}\CR \hspace{25mm} + \,[
A +\Omega + c \,,\, F + \bomega \Psi + g(\varepsilon) \eta +
i_{\varepsilon} \chi + \bomega^2 \Phi + |\varepsilon|^2 \bar\Phi] = 0
\end{gather}
$\Omega$ is the Faddeev--Popov ghost, while
$c$ is the shadow field \cite{shadow}. The action of $Q$ on the
physical fields decomposes as a gauge transformation with
parameter $c$ and a supersymmetry transformation $\susy$, as
follows \be Q= \susy -\delta^{\rm gauge }(c) \ee By dimensional
reduction of these formula one obtains the maximal subalgebra of
the $\mathcal{N}=4$ supersymmetry algebra which can be closed
off-shell \cite{N4}. By dimensional reduction in four dimensions,
the eight components of the vector $\varepsilon$ decomposes into
one scalar $\omega$, one antiselfdual $2$-form written as an
$SU(2)$ triplet $\upsilon_I$ and one four-dimensional vector
$\varepsilon^\mu$. The corresponding supersymmetry algebra is
\bea \susy A &=& \bomega \Psi + \omega \bar \Psi + g(\varepsilon)
\eta + g(J_I \varepsilon ) \chi^I + \upsilon_I J^I (\bar \Psi)\CR
\susy \Psi &=& - \bomega d_A \Phi - \omega \scal{ d_A L + T}+
i_\varepsilon F + g(J_I \varepsilon) H^I + g(\varepsilon) [\Phi,
\bar \Phi] - \upsilon_I \scal{ d_A h^I + J^I (T)}\CR \susy \Phi
&=& - \omega \bar \eta + i_\varepsilon \Psi - \upsilon_I \bar
\chi^I \CR
\susy \bar \Phi &=& \bomega \eta \CR
\susy \eta &=& \bomega [\Phi, \bar \Phi] - \omega [\bar \Phi, L] +
\Lc_\varepsilon \bar \Phi - \upsilon_I [\bar \Phi, h^I]\CR
\susy \chi^I &=& \bomega H^I + \omega [\bar \Phi, h^I] + \Lc_{J^I
 \varepsilon} \bar \Phi- \upsilon_I [\bar \Phi, L] +
{\varepsilon^I}_{JK} \upsilon^J [\bar \Phi, h^K] \CR
\susy H^I &=& \bomega [\Phi, \chi^I] + \omega \scal{ [L, \chi^I] -
 [\eta, h^I] - [\bar \Phi, \bar \chi^I]} - \Lc_{J^I
 \varepsilon} \eta - [\bar \Phi, i_{J^I \varepsilon} \Psi] +
\Lc_\varepsilon \chi^I\CR
& & \hspace{15mm} + \upsilon_J [h^J , \chi^I] + \upsilon^I \scal{
 [\eta, L] + [\bar \Phi, \bar \eta]} - {\varepsilon^I}_{JK}
\upsilon^J \scal{[\eta, h^K] + [\bar\Phi, \bar\chi^K]} \CR
\susy L &=& \bomega \bar \eta - \omega \eta + i_\varepsilon \bar
\Psi - \upsilon_I \chi^I\CR
\susy \bar \eta &=& \bomega [\Phi, L] + \omega [\Phi, \bar \Phi] +
\Lc_\varepsilon L + i_\varepsilon T + \upsilon_I \scal{ H^I + [h^I, L]}\CR
\susy \bar \Psi &=& \bomega T - \omega d_A \bar \Phi -
g(\varepsilon) [\bar \Phi, L] + g(J_I \varepsilon) [ \bar \Phi, h^I]
+ \upsilon_I J^I (d_A \bar \Phi)\CR
\susy T &=& \bomega [\Phi, \bar \Psi] + \omega \scal{-d_A \eta -
 [\bar\Phi, \Psi] + [L, \bar \Psi]} - g(\varepsilon)\scal{[\eta, L] +
 [\bar\Phi, \bar \eta]} \CR
& & \hspace{15mm} + g(J_I \varepsilon) \scal{ [\eta, h^I] +
 [\bar\Phi, \bar \chi^I]} + \Lc_\varepsilon \bar \Psi + \upsilon_I \scal{
 [h^I, \bar \Psi] + J^I (d_A \eta + [\bar\Phi, \bar \Psi])}\CR
\susy h^I &=& \bomega \bar \chi^I + \omega \chi^I - i_{J^I
 \varepsilon} \bar \Psi - \upsilon^I \eta - {\varepsilon^I}_{JK}
\upsilon^J \chi^K \\*
\susy \bar \chi^I &=& \bomega [\Phi, h^I] + \omega \scal{[L, h^I]
 - H^I} + \Lc_\varepsilon h^I - i_{J^I \varepsilon} T + \upsilon^I
[\Phi, \bar \Phi] + \upsilon_J [h^J, h^I] + {\varepsilon^I}_{JK}
\upsilon^J H^K \nonumber
\eea
Notice the presence of auxiliary fields $H_I$ and $T_\mu$, for a
total of $7=3+4$ degrees of freedom. They have been introduced to
lift some degeneracy when solving the horizontality condition,
while ensuring that the supersymmetry algebra closes, according to
\be (\susy)^2 = \delta^{\rm gauge}(\omega(\varphi) + \varpi
i_\varepsilon A) + \varpi \L_\varepsilon \label{closed}\ee
with
\be
\omega(\varphi)\equiv \bomega^2 \Phi + \bomega \omega L + \bomega
\upsilon_I h^I + (\omega^2 + \upsilon_I \upsilon^I + |\varepsilon|^2
)\bar \Phi
\ee
As explained in \cite{shadow}, the field dependent gauge
transformation that appears in the commutator of two supersymmetries
(\ref{closed}) justifies the introduction of the shadow field $c$, with the following $Q$ transformation \be Q c = \omega(\varphi) + \bomega i_\varepsilon A - c^2 \ee When
all parameters, but $\bomega$, vanish, $\omega(\varphi) $ can be
identified as a topological ghost of ghost \cite{prep}.

The $\s$ transformations of physical fields are their gauge
transformations with parameter $\Omega$.

In order to solve the degeneracy in the horizontality condition
$\s c +Q\Omega + [c, \Omega] = 0$, one introduces the field $\mu$,
with
\begin{gather}\begin{split}
\s \Omega &= - \Omega^2 \\*
\s c &= \mu \\*
\s \mu &= 0
\end{split}\hspace{10mm}\begin{split}
Q \Omega &= - \mu - [c, \Omega]\\*
%Q c &= \omega(\varphi) + \bomega i_\varepsilon A - c^2
\\*
Q \mu &= - \bigl[\omega(\varphi), \Omega\bigr] - \bomega
\Lc_\varepsilon \Omega - [c, \mu]
\end{split}\end{gather}

As explained in \cite{shadow}, the use of $c$ and $\Omega$ allows one
to disentangle the gauge symmetry and supersymmetry in the
gauge-fixing process.
In fact, antighosts and antishadows must be introduced, in order to concretely
perform a gauge-fixing, which we will choose to be $Q$-invariant. The
new fields come as a BRST quartet, and their transformation laws are
as follows:
\begin{gather}\begin{split}
\s \bar \mu &= \bar c \\* Q \bar \mu &= \bar \Omega
\end{split}\hspace{10mm}\begin{split}
\s \bar c &= 0 \\* Q \bar c &= - b
\end{split}\hspace{10mm}\begin{split}
\s \bar \Omega &= b \\ Q \bar \Omega &= \bomega \L_\varepsilon \bar \mu
\end{split}\hspace{10mm}\begin{split}
\s b &=0 \\* Q b&= - \bomega \L_\varepsilon \bar c
\end{split}\end{gather}

On all the fields of the theory one has $(d+\s+Q-\bomega
i_\varepsilon)^2=0$, that is:
\begin{gather}
 \s^2 = 0 \hspace{10mm} Q^2 = \bomega \L_\varepsilon \CR
\{\s , Q \} = 0
\end{gather}
$Q$ depends on 9 parameters. We see that if we restrict to the subalgebra with five parameters, by taking
$\varepsilon=0$, we have $Q^2=0$. This observation will be shortly used.

The grading of the fields is determined from the
assignments of the Faddeev--Popov ghost number and the shadow
number \cite{shadow}. The other quantum numbers are those for the
global symmetry $SU(2)_+\times SU(2)\times U(1)$. Together with
$\susy$, the latter invariance gives rise to a well-defined graded
subalgebra of the whole $\mathcal N=4$ symmetry, which is big enough
to completely determine the $\mathcal N=4$ action \cite{N4}. As we will see
shortly, the Ward identities associated to the invariance under this
subalgebra also determine the theory at the quantum level, in such a
way that one recovers eventually the whole symmetry of the $\mathcal{N}=4$
super-Yang--Mills theory, including its $Spin(4)\times
SL(2,\mathds{H})$ R-symmetry, after untwisting. All relevant quantum
numbers are summarized in the following tables
\pagebreak[2]
\begin{equation}
\begin{array}{|l|c|c|c|c|c|c|c|c|c|c|c|c|c|}
\hline
 & \,\, A \,\, & \,h_I\,& \,\, \Psi\,\, & \,\,\eta\,\,& \,\chi_I\,& \,\,\bar\Psi\,\,&\,\,\bar \eta\,\, &
 \,\bar\chi_I\,&\,\, \Phi\, \, &\,\,L \,\,& \,\,\bar\Phi\, \,& \,H_I\, &\,\,T\,\, \\*
\hline
{\rm canonical\,\, dimension} & 1 & 1 & \frac{3}{2} &
\frac{3}{2} & \frac{3}{2} & \frac{3}{2} & \frac{3}{2}
&\frac{3}{2}& 1&1&1&2&2 \\*
\hline
U(1) & 0 & 0 & 1&-1&-1&-1& 1&1&2&0&-2&0&0\\*
\hline
SU(2)_+ &
\frac{1}{2}&0&\frac{1}{2}&0&0&\frac{1}{2}&0&0&0&0&0&0&\frac{1}{2}\\*
\hline
SU(2) &
\frac{1}{2}&1&\frac{1}{2}&0&1&\frac{1}{2}&0&1&0&0&0&1&\frac{1}{2}\\*
\hline
{\rm commutation\ property } & -&+&+&-&-&+&-&-&+&+&+&+&-\\*
\hline
\end{array} \CR
\end{equation}\nopagebreak[4]
\begin{centerline} {\small Table I : Quantum numbers of physical
 fields (with ghost and shadow number zero)}\end{centerline}\pagebreak[3]
\begin{equation}
\begin{array}{|l|c|c|c|c|c|c|c|c|c|c|c|c|c|c|c|c|}
\hline
 & \,\, \Omega \,\, & \,\,\bar \Omega\,\,& \,\, b\,\, &
 \,\,\bar \mu\,\,& \,\,\bar c\,\,&
 \,\,c\,\,&\,\,\mu\,\, & \,\,\bomega\,\,&\,\,
 \omega\,\,&\,\,\varepsilon\,\,&\,\,\upsilon_I\,\,&\ \
 \, \, \upchi\ \ \, \,&
 \,\,\chis\,\,&\,\,\chiq\,\,&\,\,\chiqs\,\,\\*
\hline
{\rm canonical \ dimension}
&0&2&2&\frac{3}{2}&\frac{3}{2}&\frac{1}{2}&\frac{1}{2}&0&0&0&0&d&4-d&\frac{7}{2}-d&\frac{7}{2}-d\\*
\hline
{\rm ghost\,\,number} &1&-1&0&-1&0&0&1&0&0&0&0&g&-1-g&-g&-1-g\\*
\hline
{\rm shadow\,\,number} &0&0&0&-1&-1&1&1&1&1&1&1&s&-s&-1-s&-1-s\\*
\hline
U(1) & 0&0&0&0&0&0&0&-1&1&1&1&u&-u&-u&-u\\*
\hline
SU(2)_+ & 0&0&0&0&0&0&0&0&0&\frac{1}{2}&0&j_+&j_+&j_+&j_+\\*
\hline
SU(2) & 0&0&0&0&0&0&0&0&0&\frac{1}{2}&1&j&j&j&j\\*
\hline
{\rm commutation\ property } & -&-&+&+&-&-&+&+&+&+&+&\pm&\mp&\mp&\pm\\*
\hline
\end{array} \CR
\end{equation}\nopagebreak
\begin{centerline} {\small Table II : Quantum numbers of ghosts,
 shadows and sources for $\s,\, Q, \,\s Q$ transformations}
\end{centerline}\nopagebreak
where $\upchi$ stands for any field of the theory.

\subsection{Classical action and gauge-fixing}
\label{classical}
The classical action $S$ is defined as a gauge-invariant local
functional in the physical fields, which has canonical dimension four,
is invariant under the global symmetry $SU(2)_+\times SU(2) \times
U(1)$ and under the $Q$ symmetry with its nine supersymmetry
generators. The corresponding lagrangian density is any given
linear combination of
\begin{multline} \label{lag}
\L^0_4 \equiv \trace \biggl( \frac{1}{2} F_{\, \wedge}F + H_I J^I
\star F - \star H_I H^I + \varepsilon_{IJK} \star H^I[h^J,h^K] + \star
H_I[L, h^I] + T\star T \\*+ T\star d_A L + J_I \star T d_A h^I + d_A \Phi
\star d_A \bar\Phi + \chi_I J^I \star d_A \Psi + \Psi \star d_A \eta -
\bar \chi_I J^I \star d_A \bar \Psi - \bar \Psi \star d_A \bar \eta \\*+
\star \eta [\Phi, \eta] + \Psi [\bar \phi, \star \Psi] + \star \chi_I
[\Phi, \chi^I] + \star \bar \eta [\bar \Phi, \bar \eta] + \bar \Psi
[\Phi, \star \bar \Psi] + \star \bar \chi_I[\bar \Phi, \bar \chi^I] \\*-
\star \eta[L, \bar \eta] - \Psi [L, \star \bar \Psi] - \star \chi_I
[L, \bar \chi^I] - \star \eta[h_I, \bar \chi^I] + \star \bar \eta
[h_I, \chi^I] \\*+ J_I \star \Psi [h^I,
\bar \Psi] - \star \varepsilon_{IJK} \chi^I [h^J, \bar \chi^K] -
\star [\Phi, \bar \Phi]^2 - \star [\Phi, h_I] [\bar \Phi, h^I] - \star
[\Phi, L][\bar \Phi, L] \biggr)
\end{multline}
and the of topological term
\be\label{lagc}
Ch^0_4 \equiv \frac{1}{2}\trace F_{\, \wedge} F
\ee
Here, we are only interested in the sector of zero instanton number
$\int Ch^0_4 = 0$, in such a way that
\be \label{laggg} S = \frac{1}{g^2} \int \L^0_4 \ee
The only physical parameter of the theory is thus the coupling constant
$g$. Modulo the elimination of the auxiliary fields $H$ and $T$, and
the addition of a topological term, $\L^0_4$ can be identified as the
untwisted $\mathcal{N}=4$ supersymmetric lagrangian.

One can check that the action $S$ is in fact invariant under the
sixteen generators of supersymmetry. Remarkably, we found that it is
already completely determined by the $\susy$ invariance, when the
supersymmetry generators are reduced to the six ones associated to
$\bomega$, $\omega$ and $\varepsilon$ \cite{N4}.

As shown in \cite{shadow}, the introduction of the trivial BRST
quartet $(\bar \mu, \bar c, \bar \Omega ,b )$ allows a renormalizable
supersymmetric (i.e. $Q$ invariant) gauge-fixing $\s \Uppsi$ for
$S$, where $\Uppsi$ is a $Q$-exact gauge fermion that depends on the
shadow fields and on the supersymmetry parameters
\be\label{class}
 \Uppsi = Q \int \trace \bar \mu ( d \star A - \alpha b) 
\ee
Here, we will restrict to the shadow-Landau gauge
$\alpha=0$. More Ward identities exist in this gauge, which
 is stable under renormalization. They greatly simplify the renormalization problems. When $\alpha=0$, the supersymmetric
gauge-fixing action is
\begin{multline}
\s \Uppsi = \int \trace \Bigl( b d \star A - \bar \Omega d \star d_A
 \Omega + \bar c d \star d_A c + \bar \mu d \star d_A \mu \\*
 - \bar c d \star \scal{\bomega \Psi + \omega \bar \Psi + g(\varepsilon) \eta
+ g(J_I \varepsilon ) \chi^I + \upsilon_I J^I (\bar \Psi)} \\*
+ \bar \mu d \star \scal{ [d_A \Omega , c] + [\Omega ,\bomega \Psi +
 \omega \bar \Psi + g(\varepsilon) \eta + g(J_I \varepsilon ) \chi^I
 + \upsilon_I J^I (\bar \Psi)]} \Bigr)
\end{multline}

Let $\phis $, $\phiq $ and $\phiqs $ be respectively the sources
of the $\s$, $Q$ and $\s Q$ transformations of the fields. The
 gauge-fixed complete action $\Sigma$, including the
insertions of these operators, is
\begin{multline}
\Sigma = S + \s \Uppsi
+ \sum_\A \int (-1)^\A \Scal{ \phis_\A \s \varphi^\A +\phiq_\A Q \varphi^\A +
\phiqs_\A \s Q \varphi^\A}\\*
+ \int \trace \Scal{ \Omegas \Omega^2 - \Omegaq Q \Omega - \Omegaqs \s
 Q \Omega + \muq Q \mu - \cq Q c}
\end{multline}
Owing to the source dependence, the $\s$ and $Q$ invariance can be
expressed as functional identities, namely the Slavnov--Taylor
identities, defined in \cite{shadow} \be \S(\Sigma) = 0
\hspace{10mm} \Q (\Sigma) = 0 \ee Choosing the class of ``linear
gauges''~(\ref{class}), one has equations of motion for the quartet $(\bar \mu,
\bar c, \bar \Omega ,b )$ that imply functional identities
$\Gg(\Sigma)= 0$, which can be used as Ward identities. Moreover,
the shadow-Landau gauge allows for further functional
identities, associated to the equations of motion of $\Omega$, $c$
and $\mu$, $\aGg(\Sigma) = 0$, which constitute a BRST quartet
with the global gauge transformations. We refer to \cite{shadow}
for the detailed expressions of these antighost and ghost Ward
identities. All Ward identities verify consistency conditions and
their solutions determine a Lie algebra of linear functional
operators. The linearized Slavnov--Taylor operators associated to
$\s$ and $Q$ satisfy the algebra
\begin{gather}
{\S_{|\Sigma}}^2 = 0 \hspace{10mm} {\Q_{|\Sigma}}^2 = \bomega
\P_\varepsilon \CR
\bigl\{\S_{|\Sigma}, \Q_{|\Sigma}\bigr\} = 0
\end{gather}
 $\P_\varepsilon$ is the differential operator which acts as the Lie
 derivative along $\varepsilon$ on all fields and external sources.
 It must be noted that the Green functions 
 depend on the supersymmetry parameters
 generated by the $Q$-exact gauge-fixing term, but not the physical
 observables~\cite{shadow}.

\section{Renormalization of the action}
The problem of the renormalization of supersymmetric theories is
strongly simplified in the case where the supersymmetry algebra
closes without the use of the equations of motion, provided one
uses shadow fields \cite{shadow}. To treat the
renormalization of the $\mathcal N=4$ model, for which no set of
auxiliary fields exists,
 we will adapt the results of \cite{shadow}, using the maximal off-shell
closed $Q$ symmetry of the $\mathcal N=4$ model, with nine
generators. This reduces the question of the anomalies and
of the stability of the theory to algebraic problems, which involve
only the physical fields and the differential $\susy$.

\subsection{Anomalies}
The possible anomalies associated to the Ward identities \be
\S(\Gamma) = 0\hspace{10mm} \Q(\Gamma)=0 \hspace{10mm} \Gg(\Gamma)
= 0 \hspace{10mm} \aGg(\Gamma) = 0 \ee are related to the
cohomology $\mathcal{H}^\ast = \oplus_{s\in \mathds{N}} \mathcal{H}^s$
of the differential complex of gauge-invariant functionals in the
physical fields and the supersymmetry parameters ($\varpi,\, \omega,\,
\upsilon_I$ and $\varepsilon^\mu$) with differential $\susy$. The shadow number
$s$ defines the grading of this complex. The non-trivial
anomalies are either elements of $\mathcal{H}^1$ or a doublet made
 out of the Adler--Bardeen anomaly and of
a supersymmetric counterpart, which exists if and only if the
cocycle \be\label{cocycle} \int \trace \scal{ F_{\, \wedge} \susy
A_{\,\wedge} \susy A + \omega(\varphi) F_{\, \wedge} F} \ee is
$\susy$-exact \cite{shadow}.
If this cocycle were $\susy$-exact in the $\mathcal N =4$ theory, its
restriction to the value ($\varpi=1,\,\omega=\upsilon=\varepsilon=0$) of the
supersymmetry parameters would also be $\susy_{\hspace{-4mm}
 \scriptscriptstyle 1} \hspace{3mm}$-exact. This restricted operator
$\susy_{\hspace{-4mm} \scriptscriptstyle 1} \hspace{3mm}$ can be
identified with the equivariant form of the topological BRST operator
defined in the generalized Donaldson--Witten theory associated to twisted
$\mathcal{N}=4$ super-Yang--Mills theory \cite{N4} and the cocycle
(\ref{cocycle}) can be identified with the Donaldson--Witten invariant
\be \int \trace \scal{ \Phi F_{\, \wedge} F + \Psi_{\, \wedge} \Psi_{\,
 \wedge} F} \ee
The latter expression is a non-trivial cohomology class. Thus the consistency
equations for both $\s$ and $Q$ symmetries forbid the possibility of
an Adler--Bardeen anomaly in $\mathcal{N}=4$ super-Yang--Mills theory.
In fact, one obtains an analogous result for the cases $\mathcal{N}=2,3$.

This very simple demonstration gives an algebraic proof of the
absence of Adler--Bardeen anomaly in Yang--Mills theories with
extended supersymmetry.

%This very well known property is usually
%attributed to the absence of chiral fermions in the field content of
%these theories.

As for the absence of purely supersymmetric anomaly, one can
 straightforwardly compute that $\mathcal{H}^1$ is empty.
 Indeed, the invariance under only six supersymmetry generators is sufficient to
determine the classical action \cite{N4}. Then, one finds that the possible elements of
$\mathcal{H}^1$ associated to transformations linear in $\bomega$, $\omega$,
 and
$\varepsilon$ must be
trivial. After their elimination, power counting forbids the possibility
of functionals, which are linear in the parameter
$\upsilon_I$ and satisfy
 all the invariances required by the consistency
conditions.

As a corollary, one finds that, when one renormalizes the theory and adjusts
the 1PI generating functional at a given order of perturbation theory by
adding non-invariant counter-terms, it is sufficient to consider the
Slavnov--Taylor identity with the six generators. Once this is done,
it is automatic that one has also restored the Slavnov--Taylor
identity with nine generators.

Therefore, in the absence of a solution to the consistency conditions
of the functional operators associated to the Ward identities, it is by
definition possible to renormalize the $\mathcal{N}=4$ super-Yang--Mills theory, while maintaining all
the Ward identities of the shadow-Landau gauge. The process is
straightforward, and independent of the choice of the regularization.

\subsection{Stability}
In order to ensure that the renormalized action
does not depend on more parameters than the classical lagrangian that
we are starting from, one must prove its stability
property. Basically, this amounts to prove that the most general
local solution of the Ward identities, which can be imposed in
perturbation theory, has the same form as the local gauge-fixed
effective action that one starts from to define perturbation theory.

In \cite{shadow}, within the framework of our class of renormalizable gauges,
we reduced the question of the stability of the action
 to that of finding the most general
supersymmetry algebra acting on the set of physical fields of the
theory. Using power counting, we checked by inspection 
that, for the $\mathcal{N} =4$ theory, the solution of this
problem is unique, modulo a rescaling of each physical field and
modulo a redefinition of the auxiliary fields $H_I$ and $T$
\bea\label{redef} H^\R_I &=& z_{10} H_I + z_{11} J_I^{\mu\nu}
F_{\mu\nu} + z_{12} \varepsilon_{IJK} [h^I, h^K] + z_{13} [L, h_I]
\CR T^\R_\mu &=& z_{20} T_\mu + z_{21} D_\mu L + z_{22} {J_{I\,
\mu}}^\nu D_\nu h^I \eea Here the $z$'s are arbitrary
coefficients. Such a non linear renormalizations can in fact be
avoided. For this, one defines $H^I$ and $T$
 in such way that
\be \frac{\delta^L S}{\delta H^I} = -2\, \star H_I \hspace{10mm}
\frac{\delta^L S}{\delta T} = 2 \star T \ee
The auxiliary fields then decouple and are not renormalized. This property
can be checked by using Ward identities associated
to the equations of motion of these auxiliary fields, which are consistent with
the whole set of Ward identities of the theory. To define
such Ward identities, one adds to the action sources that are tensors of rank
two in the adjoint representation of the gauge group, for the local operator
$c\otimes \Omega$ (where the tensor product is for the adjoint
representation of the gauge group) and for its $\s$, $Q$ and
$\s Q$ variations.

It follows that the source independent part, $S^\R+ \s^\R\, \,
\Uppsi^\R $, of the most general local solution of Ward identities is
determined by its invariance under both renormalized symmetries
$\s^\R $ and $Q^\R$. The graded differential operators $\s^\R $ and
$Q^\R$ have the same expression as $\s $ and $Q$, with a mere
substitution of the bare fields and the coupling constant into
renormalized ones. Modulo these substitutions, $S^\R$, defined as the
most general local functional of ghost and shadow number zero, power counting four, and
invariant under all the global symmetries, which is invariant under
$\susy^\R$ and belongs to the cohomology of $\S_{|\Sigma}$, is the
same as $S$ in Eq.~(\ref{laggg}). In our class of gauge, the
gauge-fixing term keeps the same form, due to the ghost and antighost
Ward identities. One can thus write, for the most general possible
local solution of Ward identities for the $\mathcal{N}=4$ theory
\begin{multline}
\Sigma^\R = \frac{1}{{g_\R}^2} \int \L_4^{0\, \R} + \s^\R Q^\R \int \trace
\bar \mu d \star A\\*
+ \sum_\A \int (-1)^\A \Scal{ \phis_\A \s^\R \varphi^\A +\phiq_\A Q^\R
  \varphi^\A + \phiqs_\A \s^\R Q^\R \varphi^\A}\\*
+ \int \trace \Scal{ -\Omegas \s^\R \Omega - \Omegaq Q^\R \Omega -
 \Omegaqs \s^\R Q^\R \Omega + \muq Q^\R \mu - \cq Q^\R c}
\end{multline}

\subsection{Callan--Symanzik equation}\label{callan}
We define $\rp$ as the subtraction point. The renormalized
generating functional $\Gamma$ of 1PI vertices of fields and
insertion of $\s$, $Q$ and $\s Q$ transformations of all fields
verifies by construction the Callan--Symanzik equation
\be \C \, \Gamma = 0 \label{CSe}\ee
 With our choice of gauge, the supersymmetry parameters 
 do not get renormalized, because of the Ward identities.
 Thus, in the shadow-Landau gauge, the unique parameter of the
 theory that can be possibly renormalized is the coupling constant
 $g$.

Because of the quantum action principle, $\rp \frac{\partial
 \Gamma}{\partial \rp}$ is equal to the insertion
 of a local operator in the 1PI generating functional satisfying all
 the linearized functional identities associated to the Ward
 identities. Using furthermore the stability property of the effective
 action, one obtains that the anomalous dimensions of the
 fields can be adjusted, order by order in perturbation theory, in such
 a way that the Callan--Symanzik operator takes the following form
\begin{multline} \label{CS}
\C \,\F \equiv \rp \frac{\partial \F}{\partial \rp} + \beta
\frac{\partial \F}{\partial g} + \S_{|\F} \Q_{|\F} \int \biggl( \,
\sum_\A \gamma^\A \, \varphi^\A \phiqs_\A + \gamma^\ord{A} \, \trace
\bar \mu \frac{\delta^L \F}{\delta b} \biggr) \\*
= \rp \frac{\partial \F}{\partial \rp} + \beta \frac{\partial
 \F}{\partial g} - \sum_\A \gamma^\A \int \Scal{
 \varphi^\A \frac{\delta^L \F}{\delta \varphi^\A}- \phis_\A
 \frac{\delta^L \F}{\delta \phis_\A} - \phiq_\A \frac{\delta^L \F}{\delta
 \phiq_\A } - \phiqs_\A \frac{\delta^L \F}{\delta \phiqs_\A}}\\* +
\gamma^\ord{A} \int \trace \Scal{ \bar \mu
 \frac{\delta^L \F }{\delta \bar \mu} + \bar c \frac{\delta^L \F
 }{\delta \bar c} + \bar \Omega \frac{\delta^L \F }{\delta \bar
 \Omega} + b \frac{\delta^L \F }{\delta b}} \hspace{10mm}
\end{multline}
Any given local operator $\mathcal{O}_A$ generally
mixes under renormalization with all other operators with equal or
lower canonical dimensions, except if a symmetry forbids this phenomenon.

%In what follows, we
%indicate how to write the Callan--Symanzik for the general composite
%local operators.

To generate insertions of any observable $\mathcal{O}_A$ in the
1PI generating functional $\Gamma$, one couples them to external
sources $u^A$ and redefine \be \Sigma \to \Sigma[u] = \Sigma +
\sum_A \int \bigl<u^A , \mathcal{O}_A \bigr> \ee Renormalization
can only mix a finite number of local operators, because of power
counting. To control renormalization, one must generically
introduce new sources $v^X$ for other operators
and extend $\Sigma[u]$ into $\Sigma[u,v]$ in such a
way that one can define the $\s$ and $Q$
transformations of sources $u^A$ and $v^X$ so that
$\Sigma[u,v]$ satisfies all the Ward identities of
the theory. By doing so, the Slavnov--Taylor, ghost and antighost
operators get modified by source dependent terms.
Then, for any given observable with a given canonical dimension,
the theory generated by $\Sigma[u,v]$ can be renormalized
 in such a way that it satisfies the same Ward
identities as the theory generated by $\Sigma$, provided one has
introduced the large enough but finite set of sources $v^X$ and that
the introduction of these new sources does not generate anomalies.

The quantum action principle implies that the Callan--Symanzik
equation for the 1PI generating functional $\Gamma[u,v]$ can be written
as follows
\be \C \, \Gamma[u,v] = \bigl[ L[u,v] \cdot \Gamma[u,v]
\bigr] \ee
The right hand side stands for the insertion of a local functional
$L[u,v]$ of canonical dimension four in $\Gamma[u,v]$. Because of the
commutation property between the Callan--Symanzik operator and the
functional operators associated to the Ward identities of the theory,
this insertion must satisfy all the modified linearized functional identities
associated to the Ward identities including the source dependence
\begin{gather}
\S_{|\Gamma} \bigl[ L \cdot \Gamma \bigr] = 0 \hspace{10mm}
\Q_{|\Gamma} \bigl[ L \cdot \Gamma \bigr] = 0 \CR
\LGg \,\bigl[ L \cdot \Gamma \bigr] = 0 \hspace{10mm}
\LaGg \,\bigl[ L \cdot \Gamma \bigr] = 0
\end{gather}
Therefore, the local functional $L$ must be invariant under all the
global symmetries of the theory and verify
\be \S_{|\Sigma} L = \Q_{|\Sigma} L = \LGg\, L = \LaGg \,L = 0 \ee
where the linearized operators are assumed to contain the source
dependent modifications associated to $\Sigma[u,v]$.
The most general form of $L$ thus corresponds to the most general
$u,v$ dependent
invariant counterterm, solution of Ward identities~\cite{PS}.

\section{Physical observables}
Physical observables are defined as the correlation functions of
gauge-invariant functionals $\mathcal{O}^\p_A$ of
 physical fields,
\be \bigl< \mathcal{O}^\p_A \, \mathcal{O}^\p_B \,
\mathcal{O}^\p_C \cdots \bigr> \ee They belong to the cohomology
of $\s$. Thus, they do not depend on the gauge parameters,
including the supersymmetry parameters \cite{shadow}. One can
study them in the shadow-Landau gauge without loss of
generality. We mentioned in section \ref{callan} that the
supersymmetry parameters are not renormalized in this gauge.
Thus, for any set of functions of the supersymmetry parameter
$f^A(\bomega , \omega, \upsilon_I,\varepsilon)$, one has \be
\bigl< \, \Scal{ \sum_A f^A \mathcal{O}^\p_A } \, \mathcal{O}^\p_B
\, \mathcal{O}^\p_C \cdots \bigr> = \sum_A f^A \bigl<
\mathcal{O}^\p_A \, \mathcal{O}^\p_B \, \mathcal{O}^\p_C \cdots
\bigr> \label{fact}\ee The Slavnov--Taylor identities imply that
the insertion of any given gauge-invariant functional in the
physical fields $\mathcal{O}^\p_A$ are renormalized such that \be
\bigl[ \susy \mathcal{O}^\p_A \cdot \Gamma \bigr] = \Q_{|\Gamma} \bigl[
\mathcal{O}^\p_A \cdot \Gamma \bigr] \ee This equation and the
factorization property (\ref{fact}) imply that physical
observables fall into supersymmetry multiplets, when they are
sandwiched between physical states.

Eq.~(\ref{fact}) is a useful property, since it is often
convenient to introduce field functionals under the
form $\mathcal{O}= \sum_A f^A \mathcal{O}^\p_A $. One can study
the observables $\bigl< \mathcal{O}^\p_A \cdots \bigr>$ through
correlations functions involving the functional $\mathcal{O}$, as
long as each $\mathcal{O}^\p_A$ is unambiguously defined by
$\mathcal{O}$ at the classical level.

In the shadow-Landau gauge, it is therefore meaningful to define the
physical observables as the field functionals in the cohomology of
$\s$, including the ones with a dependence on the supersymmetry
parameters. Observables are allowed to have an arbitrary positive
shadow number.

\section{Protected and $1/2$ BPS operators}
Some of the local operators of the $\mathcal{N}=4$ super-Yang--Mills theory
are protected from renormalization. A strong definition of this
property is \be \C\, \bigl[ \mathcal{O} \cdot \Gamma \bigr] = 0
\ee expressing the vanishing of the corresponding
anomalous dimension, $\gamma_{\mathcal{O}}=0$. However, the form
of this equation must be slightly relaxed, since we are interested
in physical operators, and their finiteness is only
meaningful for their values between physical states. We thus
define a protected physical operator by the request that it
satisfies the previous condition, up to an unphysical $\s$-exact term,
namely
\be \C \, \bigl[ \mathcal{O}^\p \cdot \Gamma \bigr] = \S_{|\Gamma}
\bigl[ \Upsilon^\ord{\mathcal{O}} \cdot \Gamma \bigr] \label{protec}\ee
Here $\Upsilon^\ord{\mathcal{O}}$ is a local functional of ghost number
$-1$. Its expression can be gauge-dependent.

Well-known protected local operators of the $\mathcal{N}=4$ super
Yang--Mills theory are those belonging to BPS multiplets. In
superconformal theory it is natural to classify the physical
observables in irreducible superconformal multiplets. In each
superconformal multiplet, there is a superconformal primary
operator that is annihilated by the so-called special
supersymmetry generators at the point $x^\mu =0$. Moreover, the
action of supersymmetry generators on a superconformal primary operator 
generates all operators of its superconformal multiplet. When
at least one of the supercharges commutes with the superconformal
primary operator of a superconformal multiplet, the latter is
called BPS. Such irreducible multiplets are short. They play
an important role in the $AdS/CFT$ correspondence. Superconformal
invariance implies that the dimension of any operators belonging
to such multiplets do not receive radiative
corrections.

 The $1/2$
BPS primary operators are the primary operators that are annihilated by
half of the supersymmetry generators. They are the gauge-invariant
polynomials in the scalar fields of the theory in a traceless
symmetric representation of the $SO(5,1)$ R-symmetry group.
In this section we will prove that all the $1/2$ BPS
primary operators, and thus all their descendants, are protected
operators, without assuming that the theory is conformal, using only
Ward identities associated to gauge and supersymmetry invariance.

In the gauge $\varepsilon = 0$ the operator $Q$ is nilpotent. The
Lie algebra valued function of the scalar fields $\omega(\varphi)$
that characterizes the field dependent gauge transformations that
appear in the commutators of the supersymmetries
depends in this case on five parameters, \be \omega(\varphi) =
\bomega^2 \Phi + \bomega \omega L + \bomega \upsilon_I h^I +
(\omega^2 + \upsilon_I \upsilon^I )\bar \Phi \ee If we 
had considered all the supersymmetry generators,
$\omega(\varphi)$ would take the form \be
\omega(\varphi)_{\scriptscriptstyle \rm 16} =
\scal{\overline{\epsilon} \tau^i \epsilon} \phi_i \ee \
where $\tau^i$ are the six-dimensional gamma matrices for the
$SL(2,\mathds{H})$ spinor representations. We have not indicated the
$SL(2,\mathds{H})$ index for the spinor $\epsilon$. The quantity
$\omega(\varphi)$ is a particular case of
$\omega(\varphi)_{\scriptscriptstyle \rm 16}$, when $\epsilon$
satisfies, among other conditions, that
$(\overline{\epsilon}\gamma^\mu\epsilon)=0$. Thus, the expansion of
any invariant polynomial in $\omega(\varphi)$ in powers of the
supersymmetry parameters gives operators that belong to symmetric
representations of $SO(5,1)$. Moreover, because of the original
ten-dimensional Fiertz identity \be \scal{\overline{\upepsilon} \Gamma_m
\upepsilon} \Gamma^m \upepsilon = 0 \ee for any given commuting
Majorana--Weyl spinor $\upepsilon$, any given $\mathcal{N}=4$
Majorana spinor that satisfies
$(\overline{\epsilon}\gamma^\mu\epsilon)=0$, is such that \be
\scal{\overline{\epsilon} \tau^i \epsilon}
\scal{\overline{\epsilon} \tau_i \epsilon}=0 \ee Therefore, all
operators obtained from the expansion of $\omega(\varphi)$ belong to
traceless symmetric representations of $SO(5,1)$.

In fact, the invariant polynomials $\mathcal{P}\scal{
\omega(\varphi)}$ give by expansion in the supersymmetry parameters
the whole traceless symmetric representations of $SO(5,1)$. To
obtain this result, it is sufficient to show that this expansion
provides an equal number of operators than there are components in
the representations. It is convenient to
use a four dimensional notation, with $\gimel=(0,I)$, $\upsilon^\gimel
\equiv (\omega, \upsilon^I)$ and $h_\gimel \equiv (L,
h_I)$. Call $X(n_+,n_-)$ the sum of the monomials, of degree $n_+$ in $\bomega$
and $n_-$ in $\upsilon^\gimel$, which may stand in the expansion of
$\mathcal{P}\scal{\omega(\varphi)}$. For $n_+ \geqslant n_-$,
$X(n_+,n_-)$ takes the following form\footnote{For simplicity we have
 written $X(n_+,n_-)$ in the simplest case where
 $\mathcal{P}\scal{\omega(\varphi)}= \trace
 \omega(\varphi)^{\frac{n_++n_-}{2}}$. The demonstration extends
 trivially to any invariant polynomials.}
\be
X(n_+,n_-) \propto {\rm s}\trace
\Phi^{\frac{n_+-n_-}{2}} \Scal{
 (\upsilon^\gimel h_\gimel)^{n_-}+ \sum_{p=1}^{\frac{n_-}{2}} C^p_{n_+
   \, n_-} (\upsilon^\gimel h_\gimel)^{n_-  -2p}(\upsilon_\gimel
 \upsilon^\gimel \Phi \bar \Phi)^p} 
\ee
where ${\rm s}\trace$ is the symmetrized trace and $C^p_{n_+\, n_-}=
\frac{n_- ! (\frac{n_+ - n_-}{2})!}{p!(n_- - 2 p)!(p+ \frac{n_+ -
    n_-}{2})!}$. By defining $S_d^n =\frac{(d+n-1)!}{(d-1)!n!}$ as the
dimension of the symmetric representation of rank $n$ in $SO(d)$,
$X(n_+,n_-)$ gives $S_4^{n_-}$ operators. For $n_+< n_-$, $X(n_+,n_-)$
takes the form
\be
X(n_+,n_-) \propto {\rm s}\trace
(\upsilon_\gimel \upsilon^\gimel \bar \Phi)^{\frac{n_- - n_+}{2}} \Scal{
 (\upsilon^\gimel h_\gimel)^{n_+}+ \sum_{p=1}^{\frac{n_+}{2}}
 C^p_{n_-\, n_+}  (\upsilon^\gimel h_\gimel)^{n_+
   -2p}(\upsilon_\gimel \upsilon^\gimel \Phi \bar \Phi)^p}
\ee
and gives $S_4^{n_+}$ operators. By expanding an
invariant polynomial of degree $n$ as a power series in the
supersymmetry parameters, one thus obtains \be\sum_{n_-=0}^{n} S_4^{n_-}
+ \sum_{n_+=0}^{n-1} S_4^{n_+}= S_4^n + 2 \sum_{p=0}^{n-1}
S_4^p \ee independent operators in the traceless symmetric representation
of $SO(5,1)$. The traceless symmetric representation of $SO(5,1)$ of
rank $n$ is of dimension $S_6^n - S_6^{n-2}$. One can then compute by 
recurrence that
\be S_{d-2}^n + 2 \sum_{p=0}^{n-1} S_{d-2}^p =
S_d^n - S_d^{n-2} \ee One has \be S_{d-2}^2 + 2 ( S_{d-2}^0 +
S_{d-2}^1) = S_d^2 - S_d^0 = \frac{(d-1)(d+2)}{2} \ee and \bea
S_{d-2}^n + 2 \sum_{p=0}^{n-1} S_{d-2}^p - \scal{ S_{d-2}^{n-1} +
2
 \sum_{p=0}^{n-2} S_{d-2}^p} &=& S_{d-2}^n + S_{d-2}^{n-1} \CR
S_d^n - S_d^{n-2} - \scal{ S_d^{n-1} - S_d^{n-3}} &=&
\frac{(d+n-4)!}{(d-1)!n!} (2n + d-3) \eea 

Thus, we finally have
the result that any gauge invariant polynomial in the scalar
fields that belongs to a traceless symmetric
representations of $SO(5,1)$ can be represented by an invariant
polynomial $\mathcal{P}$ in $\omega(\varphi)$.

Since  $Q^2=0$ with the restricted symmetry with   the five parameters $\omega,\bomega,
v^I$, we can use     the horizontality condition
(\ref{extcurvature}) and the   Chern--Simons formula. It 
implies that, for any given invariant symmetrical polynomial
$\mathcal{P}$, one has
\be \mathcal{P}\scal{\omega(\varphi)} = Q \,\Delta\scal{c,\omega(\varphi)} \ee
with
\be \Delta\scal{c, \omega(\varphi)} = \int_0^1 dt\,
\mathcal{P}\scal{c\,|\, t \omega(\varphi) + (t^2- t) c^2\,} \ee
where $\mathcal{P}(X) \equiv \mathcal{P}(X,X,X,\cdots)$ and
\be
\mathcal{P}(Y|X) \equiv \mathcal{P}(Y,X,X,\cdots) +
\mathcal{P}(X,Y,X,\cdots) + \mathcal{P}(X,X,Y,\cdots) + \cdots \ee

%Because the traceless symmetric representations of $SO(5,1)$ are
%irreducible, one can deduce that

Any given polynomial in the scalar fields belonging to a traceless
symmetric representation of $SO(5,1)$ has a canonical dimension
which is strictly lower than that of all other
operators in the same representation, made out of
other fields. Thus, by
 power counting, the polynomials in the scalar fields can
 only mix between themselves under renormalization. That
is, for any homogeneous polynomial $\mathcal{P}_A$ of degree $n$ in
the traceless symmetric representation, one has \be \C \bigl[
\mathcal{P}_A\scal{\omega(\varphi)} \cdot \Gamma \bigr] = \sum_B
{\gamma_A}^B \bigl[ \mathcal{P}_B \scal{\omega(\varphi)} \cdot
\Gamma \bigr] \ee Then, the Slavnov--Taylor identities imply \be
\C \bigl[ \Delta_A\scal{c,\omega(\varphi)} \cdot \Gamma \bigr] =
\sum_B {\gamma_A}^B \bigl[ \Delta_B\scal{c,\omega(\varphi)} \cdot
\Gamma \bigr] + \cdots \ee \nopagebreak[3] where the dots stand for possible
$\Q_{|\Gamma}$-invariant corrections. However, in the shadow-Landau
gauge, $\Delta_A(c,\omega(\varphi))$ cannot appear in the right
hand side because such term would break the ghost Ward
identities. One thus gets the result that ${\gamma_A}^B = 0$
\be \C \bigl[ \mathcal{P}_A\scal{\omega(\varphi)} \cdot \Gamma \bigr] = 0 \ee
 Upon decomposition of this equation in function of the five
independent supersymmetry parameters, one then gets the finiteness
proof for each invariant polynomial
$\mathcal{P}(\phi)\equiv \mathcal{P}(\phi^i,\phi^j,\phi^k,\cdots)$
in the traceless symmetric representation of the R-symmetry group,
 namely \be \C \bigl[ \mathcal{P}(\phi) \cdot \Gamma
\bigr] = 0 \label{BPS}\ee

Having proved that all $1/2$ BPS
primary operators have zero anomalous dimension,
the $Q$-symmetry implies that
 all the operators generated from them, by applying 
 $\mathcal{N}=4$ super-Poincar\'e generators, have also vanishing
 anomalous dimensions. It follows that all the operators of
the $1/2$ BPS multiplets are protected operators.

It is worth considering as an example the simplest case of
$\trace \omega(\varphi)^2$. One has
\begin{gather}
Q \trace \scal{\omega(\varphi) c - \frac{1}{3} c^3} = \trace
\omega(\varphi)^2 \hspace{10mm} \s Q \trace \scal{\omega(\varphi) c -
 \frac{1}{3} c^3} = 0 \CR
\s \trace \scal{\omega(\varphi) c - \frac{1}{3} c^3} = \trace \Scal{
 \mu\scal{ \omega(\varphi) - c^2 } - [\Omega, \omega(\varphi)] c}
\end{gather}
Following \cite{2Sor}, one couples these operators to the theory by
adding source terms to the effective action $\Sigma$
\begin{gather}
u \trace \scal{\omega(\varphi) c - \frac{1}{3} c^3} + u^\a \trace \Scal{
 \mu\scal{ \omega(\varphi) - c^2 } - [\Omega, \omega(\varphi)] c} +
u^\q \trace \omega(\varphi)^2 \label{addi}
\end{gather}
This action satisfies the Slavnov--Taylor identities associated to
the $\s$ and $Q$ symmetries, provided that the sources $u^\bullet$
transform as follows
\begin{gather}\begin{split}
\s u^\q &= 0 \\*
\s u^\a &= u \\*
\s u \ &= 0
\end{split}\hspace{10mm}\begin{split}
 Q u^\q &= u \\*
Q u^\a &= 0 \\*
Q u\ &= 0
\end{split}\end{gather}
It is easy to check by inspection that the introduction of these
new sources cannot introduce any potential anomaly for
the Slavnov--Taylor identities. In the shadow-Landau gauge, the
ghost Ward identities remain valid, with an additional dependence
in the sources $u^\bullet$ \footnote{ For invariant polynomials
$\mathcal{P}$ of rank higher than 2, one has to
introduce further sources in order to restore the ghost Ward
identities \cite{Aless}. But we can always carry out this with a finite number
of such sources.}
\bea \int \Scal{ \frac{\delta^L
 \Gamma}{\delta \mu} - \Bigl[ \bar \mu, \frac{\delta^L
 \Gamma}{\delta b} \Bigr] + u^\a \frac{\delta^L \Gamma}{\delta \cq} -
(-1)^\A [\phiqs_\A, \varphi^\A] +
 [\Omegaqs, \Omega] + [\muq, c] } &=& 0 \CR
\int \biggl( \frac{\delta^L
 \Gamma}{\delta c} + \Bigl[\bar c, \frac{\delta^L \Gamma}{\delta b} \Bigr]-
\Bigl[ \bar \mu, \frac{\delta^L \Gamma}{\delta \bar \Omega} \Bigr]+
(-1)^\A \Bigl[
\phiqs_\A, \frac{\delta^L \Gamma}{\delta \phis_\A}\Bigr] - \Bigl[
\Omegaqs, \frac{\delta^L \Gamma}{\delta \Omegas}\Bigr] \hspace{10mm}& &\CR
+ u \frac{\delta^L\Gamma}{\delta \cq} + u^\a
 \frac{\delta^L\Gamma}{\delta \muq} + [\phiq_\A, \varphi^\A] +
 [\Omegaq, \Omega] + [\cq, c] +
 [\muq, \mu]\biggr) &=& 0 \CR
\int \biggl( \frac{\delta^L
 \Gamma}{\delta \Omega} - \Bigl[ \bar \Omega, \frac{\delta^L
 \Gamma}{\delta b}\Bigr]+ \Bigl[ \bar \mu, \frac{\delta^L \Gamma}{\delta \bar
 c}\Bigr] - \Bigl[ c, \frac{\delta^L \Gamma}{\delta \mu} \Bigr]-
(-1)^\A \Bigl[ \phiqs_\A,
\frac{\delta^L \Gamma}{\delta \phiq_\A}\Bigr] \hspace{10mm} & &\\*
 + \Bigl[ \Omegaqs, \frac{\delta^L \Gamma}{\delta
 \Omegaq}\Bigr] + \Bigl[ \muq,\frac{\delta^L \Gamma}{\delta \cq}
\Bigr]+ [\phis_\A, \varphi^\A] + [\Omegas, \Omega] \biggr) &=& 0
\nonumber \label{ghost} \eea

The most general $u^\bullet$ dependent counter term which
satisfies both Slavnov--Taylor identities is \be u^\q
\Q_{|\Sigma} \Delta^\gra{0}{3}_{[\frac{3}{2}]} + u
\Delta^\gra{0}{3}_{[\frac{3}{2}]} + u^\a \S_{|\Sigma}
\Delta^\gra{0}{3}_{[\frac{3}{2}]}\ee where
$\Delta^\gra{0}{3}_{[\frac{3}{2}]}$ must be a local functional of
ghost and shadow number $(0,3)$, canonical dimension
$\frac{3}{2}$, which verifies \be \S_{|\Sigma} \Q_{|\Sigma}
\Delta^\gra{0}{3}_{[\frac{3}{2}]} = 0 \ee
 $\Delta^\gra{0}{3}_{[\frac{3}{2}]}$ is also
 a scalar under the action of the symmetry group $SU(2)_+
\times SU(2) \times U(1)$.

These constraints imply that 
 $\Delta^\gra{0}{3}_{[\frac{3}{2}]}$ is proportional to $\trace
 \scal{\omega(\varphi) c - \frac{1}{3} c^3}$. Thus the three
 insertions that we have introduced can only be multiplicatively
 renormalized, having the same anomalous dimension.
 Moreover, the ghost Ward identities forbid the introduction of
 any invariant counter term including the shadow field $c$, if it is
 not trough a derivative term $d c$ or particular combinations of
 $c$ and the other fields that do not appear in the insertion
 $\trace \scal{\omega(\varphi) c - \frac{1}{3} c^3}$.
 This gives the result that \be \C\, \bigl[ \trace \omega(\varphi)^2 \, \cdot
\Gamma \bigr] = 0 \ee

Finally, owing to the factorization property we obtain that all
the 20 operators that constitute the traceless symmetric tensor
representation of rank two in $SO(5,1)$ are protected
operators
\begin{gather}
\trace\scal{ \Phi^2}\, , \ \trace\scal{ \Phi L} \, , \
\trace\scal{\Phi \bar \Phi + \frac{1}{2} L^2}\, ,\ \trace \scal{\bar \Phi L}\,
,\ \trace \scal{\bar \Phi^2 }\,,\label{sst}\CR
\trace\scal{ \Phi h_I} \, , \ \trace\scal{ L h_I }\, , \ \trace\scal{
 \bar\Phi h_I} \, , \ \trace\scal{ \delta_{IJ} \Phi \bar \Phi +
 \frac{1}{2} h_I h_J}
\end{gather}

This constitutes the simplest example of (\ref{BPS}), for $\mathcal{P}(\phi)
\equiv \trace \scal{ \phi^i \phi_j -\frac{1}{6}
 \delta^{i}_{j} \phi_k \phi^k }$.

\section{Cancellation of the $\beta$ function}
We now give an improved version of the proof given in
\cite{gSor}, that the $\beta$ function is zero to all order in the 
perturbative $\mathcal{N}=4$ super-Yang--Mills theory.

The proof of the cancellation of the $\beta$ function is a corollary of the three following propositions
\bea
&\bullet& \hspace{8mm} {\rm The\ } \beta{\rm\ function\ is\ zero\
 at\ first\ order\ } \label{beta1}\\*
&\bullet& \hspace{10mm} \C \ \bigl[\int \L_4^0 \cdot
 \Gamma \bigr] = \ins{1} \label{zeroZ}\\*
&\bullet& \hspace{10mm} \frac{\partial \Gamma}{\partial g} +
 \frac{2}{g^3} a(g) \, \bigl[\int \L^0_4 \cdot \Gamma \bigr] =
 \ins{2} \label{QAP}
\eea where $\Uppsi^\ord{\xi}$ are integrated insertions of ghost
number -1 and shadow number 0. Moreover, \be a(g) = 1 +
\sum_{\mathds{N}*} a_n \,g^{2n} \ee is a function of the coupling
constant $g$ that accounts for the possible radiative
corrections to the classical equation \be \frac{\partial
\Sigma}{\partial g} + \frac{2}{g^3} \, \int \L^0_4 = 0
\label{QAPc}\ee
The key part of of the proof follows
from the equation \be \Bigl[\C, \frac{\partial\,}{\partial
g}\Bigr] \F = - \frac{\partial
 \beta}{\partial g} \, \frac{\partial \F}{\partial g} - \S_{|\F}
\Q_{|\F} \int \biggl( \, \sum_\A \frac{\partial
\gamma^\A}{\partial g} \, \varphi^\A \phiqs_\A + \frac{\partial
\gamma^\ord{A}}{\partial g} \trace \bar \mu \frac{\delta^L
\F}{\delta b} \biggr) \ee If one equates $\F$ to the 1PI
generating functional $\Gamma$, one obtains, as a direct
consequence of (\ref{CSe}), (\ref{zeroZ}) and (\ref{QAP}), that
\be \frac{\partial\,}{\partial g} \Scal{\beta \, \frac{2}{g^3}
a(g) }\, \bigl[\int \L_4^0 \cdot \Gamma \bigr] = \ins{3} \ee Since
$ \bigl[\int \L_4^0 \cdot \Gamma \bigr]\neq 0 $ 
belongs to the cohomology of $\S_{|\Gamma}$, the right-hand side
and left-hand side of this equation must be zero. This implies
\be \frac{\partial\,}{\partial g} \Scal{\beta \, \frac{2}{g^3}
a(g)} \label{diffBeta} = 0 \ee This equation can be expanded in
power of $g$, $\beta = \sum_{n\in
 \mathds{N}} \beta_\ord{n} \, g^{2n+1} $. It gives
\be \beta_\ord{n} = - \sum_{p=1}^{n-1} a_{n-p}\, \beta_\ord{p}\,
\propto \beta_\ord{1},\ \ \ \ n \geqslant 2 \ee Using then the
proposition (\ref{beta1}), that is $\beta_\ord{1}=0$, one
 obtains that the $\beta$ function is zero at all
 orders in perturbation theory.

 Let us now demonstrate the basic ingredients (\ref{zeroZ},\ref{QAP})
 of the proof that the $\beta$ function vanishes to all orders.

 The cancellation of the one-loop $\beta$ function
(\ref{beta1}) is a well-established result in perturbation
theory.

 The proposition (\ref{QAP}) is a straightforward
consequence of the property that the $\mathcal{N}=4$
super-Yang--Mills action has only one physical parameter, namely
the coupling constant $g$. The Slavnov--Taylor operators commute
with the derivation with respect to $g$, and thus \be
\S_{|\Gamma} \frac{\partial \Gamma}{\partial g} = \Q_{|\Gamma}
\frac{\partial \Gamma}{\partial g} = 0 \ee Starting from the
classical equation (\ref{QAPc}),
 the quantum action principle implies that differentiation of the 1PI
generating functional with respect to the coupling
constant $g$ amounts to the insertion of an integrated local
functional of ghost and shadow number zero, which satisfies all
the global linear symmetries of the theory, and which is invariant
under the action of the two linearized Slavnov--Taylor operators
$\S_{|\Sigma}$ and $\Q_{|\Sigma}$ (see \cite{gSor} for more
details). The only such functional in the cohomology of
$\S_{|\Sigma}$ is the classical action $\int \L^0_4$, what
establishes the result (\ref{QAP}).

The only non-trivial point for proving the vanishing of the $\beta$
function is thus the demonstration of (\ref{zeroZ}), which we now
show, in the shadow-Landau gauge.

\subsection{Cocycles and descent equations for the lagrangian density}
To prove (\ref{zeroZ}), we will use the fact that the
lagrangian density is uniquely linked to a protected operator by
descent equations, involving the equivariant part of the $Q$ symmetry.

Because $\L^0_4 $ and $Ch^0_4 $ (\ref{lag},\ref{lagc}) are
supersymmetric invariant only modulo a boundary term, the
algebraic Poincar\'e lemma predicts series of cocycles, which are
linked to $\L^0_4 $ and $Ch^0_4 $ by descent equations, as follows:
\be\begin{split}
\susy \L^0_4 + d \L^1_3 &= 0 \\*
\susy \L^1_3 + d \L^2_2 &= \bomega i_\varepsilon \L^0_4 \\*
\susy \L^2_2 + d \L^3_1 &= \bomega i_\varepsilon \L^1_3 \\*
\susy \L^3_1 + d \L^4_0 &= \bomega i_\varepsilon \L^2_2 \\*
\susy \L^4_0 &= \bomega i_\varepsilon \L^3_1
\end{split}\hspace{20mm}\label{cocy}\begin{split}
\susy Ch^0_4 + d Ch^1_3 &= 0 \\*
\susy Ch^1_3 + d Ch^2_2 &= \bomega i_\varepsilon Ch^0_4 \\*
\susy Ch^2_2 + d Ch^3_1 &= \bomega i_\varepsilon Ch^1_3 \\*
\susy Ch^3_1 + d Ch^4_0 &= \bomega i_\varepsilon Ch^2_2 \\*
\susy Ch^4_0 &= \bomega i_\varepsilon Ch^3_1
\end{split}\ee
Using the grading properties of the shadow number and the form degree,
we conveniently define
\begin{gather}\L \equiv \L_4^0 + \L_3^1 + \L_2^2 + \L_1^3 + \L_0^4\CR
 Ch  \equiv  Ch_4^0 + Ch_3^1 + Ch_2^2 + Ch_1^3 + Ch_0^4 \end{gather}
The descent equations can then be written in a unified way \be (d +
\susy - \bomega i_\varepsilon ) \L = 0 \hspace{10mm}(d +
\susy - \bomega i_\varepsilon ) Ch = 0 \ee Note that on
gauge-invariant polynomials in the physical fields, $\susy$ can be
identified to $\s + Q$, in such way that the differential $(d +
\susy - \bomega i_\varepsilon )$ is nilpotent on them. Since
$\L^0_4 $ and $Ch^0_4 $ are the unique solutions of the first
equation in (\ref{cocy}), one obtains that $\L$ and
$Ch$ are the only non-trivial solutions of the descent equations,
that is, the only ones that cannot be written as
$(d+\susy-\bomega i_\varepsilon)\, \Xi$. In fact $Ch^0_4 $ and $\L^0_4
$ are the unique non-trivial solutions of Eq.~(\ref{cocy}), even when
$\susy$ is restricted to six supersymmetry parameters
($\upsilon_I=0$).

The expression of the cocycles $Ch^{s}_{4-s}$ can be simply obtained, by
changing $F$ into the extended curvature (\ref{extcurvature}) in the
topological term $\frac{1}{2} \trace FF$, owing to the horizontality equation that expresses $\s$ and $Q$.
Therefore:
\be Ch = \frac{1}{2} \trace \Bigl( F + \varpi \Psi + \omega \bar \Psi
+g(\varepsilon) \eta + g(J_I \varepsilon) \chi^I + \varpi^2 \Phi +
\varpi \omega L + (\omega^2 + |\varepsilon|^2) \bar\Phi \Bigr)^2 \ee

As for determining the explicit form of $\L^{s}_{4-s}$ for $s
\geqslant 1$, we found no other way than doing a brute force
computation, starting from $\L^0_4 $ in Eq.~(\ref{lag}). In this way,
one gets, in a step by step computation
\begin{multline}
\L^1_3 = \trace \biggl( \bomega \scal{\Psi_{\, \wedge} F + J_I ( \bar
 \chi^I T - \bar \Psi [\Phi, h^I])}\\* + \omega \scal{ \bar
 \Psi_{\, \wedge} F + J_I ( - \eta d_A h^I + \chi_I T + \Psi [\bar
 \Phi, h^I] - \bar \Psi [L, h^I])} \\*
+ J_{I\, \wedge} i_\varepsilon F \chi^I + i_\varepsilon \star \chi_I
H^I + \varepsilon_{IJK} g(\varepsilon) J^I \chi^J H^K \\*+ (i_{J^I
 \varepsilon} \star \eta - i_\varepsilon \star \chi^I) \Bigl(
 \frac{1}{2} \varepsilon_{IJK} [h^J,h^K] + [L, h_I]\Bigr) \\*
+ \star \bar \Psi i_\varepsilon T - i_\varepsilon (\bar \Psi \star d_A
L) + i_\varepsilon (J_{I\, \wedge} \bar\Psi)_{\, \wedge} d_A h^I \\*+
i_\varepsilon \star \bar \eta [\bar \Phi, L] - i_{J_I \varepsilon}
\star \bar \eta [\bar \Phi, h^I] + i_{\varepsilon} \star \eta [\Phi,
\bar \Phi] + i_\varepsilon (\Psi \star d_A \bar \Phi) - g(\varepsilon)
\Psi d_A \bar \Phi \biggr)
\end{multline}

\begin{multline}
\L^2_2 = \trace \biggl( \bomega^2 \scal{ \Phi F + \frac{1}{2} \Psi
 \Psi} + \bomega \omega \scal{ L F + \Psi \bar \Psi + J_I(h^I
 [\Phi, \bar \Phi] - \eta \bar \chi^I)} \\*
+ \omega^2 \scal{ \bar \Phi F + \frac{1}{2} \bar \Psi \bar \Psi
 + J_I (L[\bar \Phi, h^I] - \eta \chi^I)} + \bomega \scal{ g(J_I
 \varepsilon) \Psi \chi^I + i_\varepsilon J_{I\, \wedge} \bar \Psi
 \bar \chi^I - g(\varepsilon) \Phi d_A \bar \Phi} \\*
+ \omega \scal{ - g(J_I \varepsilon) \bar \Phi d_A h^I + J_I
 i_\varepsilon \bar \Psi \chi^I - 2 \eta (g(\varepsilon) \bar \Psi)^-
 - g(\varepsilon) L d_A \bar \Phi} \\*
+ \frac{1}{2} g(J_I\varepsilon) g(J_J \varepsilon) \bar \Phi [h^I,
h^J] + \star g(\varepsilon) g(J_I \varepsilon) L [\bar \Phi, h^I] +
\frac{1}{2} |\varepsilon|^2 \bar\Psi \bar \Psi - J_I i_\varepsilon
\bar \Psi i_{J^I \varepsilon} \bar \Psi \biggr)
\end{multline}

\begin{multline}
\L^3_1 = \trace \biggl( \bomega^3 \Phi \Psi + \bomega^2 \omega (L
\Psi + \Phi \bar \Psi) + \bomega \omega^2 (L \bar \Psi + \bar \Phi
\Psi) + \omega^3 \bar \Phi \bar \Psi \\*
+ \bomega^2 g(J_I \varepsilon) \Phi \chi^I + \bomega \omega
g(J_I\varepsilon) (L \chi^I -\bar \Phi \bar \chi^I) \\* + \omega^2
g(\varepsilon) \bar \Phi \eta + \omega \bar \Phi
g(J_I\varepsilon) i_{J^I \varepsilon} \bar \Psi \biggr)
\end{multline}

\be \L^4_0 = \frac{1}{2} \trace \Scal{\scal{ \bomega^2
 \Phi + \bomega \omega L + \omega^2 \bar \Phi}^2 +
 \bomega^2 |\varepsilon|^2 \bar \Phi^2 } \ee

\subsection{Finiteness property for the classical action $\int \L^0_4 $}

 The last cocycle $\L^4_0 $ is a linear combination of protected operators
 associated to the second Chern class $\trace \omega(\varphi)^2$.
 Therefore, its anomalous dimension is zero. We now show that this implies
 that its ascendant $\L^0_4 $ is also protected modulo a $d$ variation, which
 is the non-trivial condition (\ref{zeroZ}) for proving the vanishing
 of the $\beta$ function, which we now rewrite
 \be \label{m} \C \ \bigl[\int \L_4^0 \cdot
 \Gamma \bigr] = \S_{|\Gamma} \bigl[ \int \Upsilon_4^\gra{-1}{0}
 \, \cdot \Gamma ] \ee
It is worth going into the details of the proof of this identity.

To define one insertion of the lagrangian density $\L_4^0$ and its
descendants $\L^{p}_{4-p}$ in a way that preserves the
Slavnov--Taylor identities, one introduces sources for each term
$\L^{p}_{4-p}$ and defines the effective action \be \Sigma\to
\Sigma[u] \equiv \Sigma + \int \sum_p u^{-p}_p \L^p_{4-p}
\label{prime}\ee This effective action verifies the
Slavnov--Taylor identities, provided that
 the sources $u_p^{-p}$ are $\s$-invariant and
\be Q u_p^{-p} = - d u_{p-1}^{1-p} + \bomega i_\varepsilon
u_{p+1}^{-1-p} \ee
Using the extended form $u \equiv \sum_p u_p^{-p}$, one has
\be (d + Q - \bomega i_\varepsilon) u = 0 \ee

 Let us define as $\Delta^p_{4-p}$ the local
counterterms that might occur for the renormalization of the
operators $\L^{p}_{4-p}$ . The question is that of
determining the most general invariant counterterm for the
effective action $ \int \sum_p u^{-p}_p \Delta^p_{4-p}$, which is
linear in the $u_p^{-p}$. It ought to be
invariant under $SU(2)_+\times SU(2) \times U(1)$ and to obey the
Slavnov--Taylor identities: \be \S_{|\Sigma} \int \sum_p u^{-p}_p
\Delta^p_{4-p} = 0 \hspace{10mm}\Q_{|\Sigma} \int \sum_p u^{-p}_p
\Delta^p_{4-p} = 0 \label{counter}\ee If we call $\Delta \equiv
\sum_p \Delta^p_{4-p}$, one must have \be (d + \S_{|\Sigma} +
\Q_{|\Sigma} - \bomega i_\varepsilon ) \Delta = 0 \ee Since the
cohomology of $\S_{|\Sigma}$ modulo $d$ can be identified
 with the gauge-invariant polynomials in the physical
fields, $\Delta$ must be the sum of a gauge-invariant
polynomial in the physical fields and of a
$\S_{|\Sigma}$-exact term $\S_{|\Sigma} \Upsilon $, where
$\Upsilon$ is an arbitrary extended form in the fields and the
sources of ghost number $-1$. Because $\L$ and $Ch$
generate the only non-trivial elements of the cohomology of
$d+\susy-\bomega{i_\varepsilon}$ in the set of gauge-invariant
polynomials in the physical fields, $\Delta$ must be of the form
\be \Delta = z_1 \L + z_2 Ch +(d + \susy - \bomega i_\varepsilon)
\, \Xi + \S_{|\Sigma} \Upsilon \label{gen} \ee $\Xi$ is an
arbitrary gauge-invariant extended form in the physical fields of
total degree 3 and canonical dimension $\frac{7}{2}$.

We have seen in the previous section that $\L_0^4 $ is a
protected operator, and therefore, $ \C \, \bigl[ \L_0^4 \cdot
\Gamma \bigr] = 0 $. Thus, all invariant counterterms that
 might be generated in perturbation theory have to be
such that \be \Delta^4_0 = 0 \ee and therefore \be z_1 \L^4_0 +z_2
Ch^4_0 + \susy \, \Xi^3_0 - \bomega i_\varepsilon \, \Xi^2_1 = 0
\ee Each term in this expansion must separately vanish. Indeed,
$\L^4_0$ and $Ch^4_0$ are not $\susy$-exact and \be Ch^4_0 -
\L^4_0 = i_\varepsilon \, g(\varepsilon) \bar \Phi \scal{\bomega^2
\Phi + \bomega \omega L + \frac{1}{2} ( \omega^2 +
 |\varepsilon|^2) \bar\Phi} \ee
 cannot be written as a contraction with respect to the vector
$\bomega \varepsilon$ of a $1$-form that is analytic in $\bomega$.

It follows that the most general functional (\ref{gen}) which has
vanishing component of shadow number four, must have a component
of zero shadow number of the following form \be \Delta_4^0 = d \,
\Xi_3^0 + \S_{|\Sigma} \Upsilon_4^\gra{-1}{0} \ee This is
precisely the result (\ref{m}) that we wanted to prove.

\section{Conclusion}

 A great
 improvement due to the introduction of the shadow fields is
that one has two separated and consistent
Slavnov--Taylor identities corresponding respectively
to gauge and supersymmetry invariance. This enables one to
 establish the cancellation of the anomalous dimension
of some operators, considering them as insertions in any Green
functions of physical observables, which are not restricted to be
supersymmetric scalars. As for the physics, it is now defined to
be the cohomology of $\S_{|\Gamma}$ rather than the cohomology of
$\Q_{|\Gamma}$, and its supersymmetry covariance is easy to check.

Aspects of the superconformal invariance of the
$\mathcal N=4$ theory can be checked at any given finite order in
perturbation theory, for any type of ultraviolet regularization. In
this paper we have proved the cancellation of the $\beta$ function and
the finiteness of the $1/2$ BPS operators. The method can certainly be
extended to other features of superconformal invariance.

\subsection*{Acknowledgments}

This work was partially supported under the contract ANR(CNRS-USAR) \\ \texttt{no.05-BLAN-0079-01}.

%%%%%%%%%%%%%%%%%%%%%%%%%%%%%%%%%%%%%%%%%%%%%%%%%%%%%%%%%%%%%%%%%%%%%%%%%%%%%

\end{document}